\documentclass[12pt]{iopart}
\usepackage{times}
\usepackage{graphicx}
\usepackage{amsfonts}
\usepackage{amssymb}
\usepackage{amsbsy}

\newcommand{\BEQ}{\begin{equation}}     
\newcommand{\BEA}{\begin{eqnarray}}
\newcommand{\EEQ}{\end{equation}}       
\newcommand{\EEA}{\end{eqnarray}}

\renewcommand{\vec}[1]{\boldsymbol{#1}} 

\newcommand{\vekz}[2]
     {\mbox{${\begin{array}{c} #1  \\ #2 \end{array}}$}}

\begin{document}

\title{Local scale invariance, conformal invariance and dynamical scaling}
\author{Malte Henkel}
\address{Laboratoire de Physique des Mat\'eriaux CNRS UMR 7556, Universit\'e
Henri Poincar\'e Nancy~I, B.P. 239, F - 54506 Vand{\oe}uvre l\`es Nancy Cedex, 
France}
\begin{abstract}
Building on an analogy with conformal invariance, local scale transformations 
consistent with dynamical scaling are constructed. 
Two types of local scale invariance are found which act as dynamical 
space-time symmetries of certain non-local
free field theories. Physical applications include uniaxial Lifshitz points and 
ageing in simple ferromagnets. 
\end{abstract}

Scale invariance is a central notion of modern theories of critical and
collective phenomena. We are interested in systems with strongly anisotropic
or dynamical criticality. In these systems, two-point functions satisfy the
scaling form
\begin{equation} \label{gl:skala}
G(t,\vec{r}) = b^{2x} G(b^{\theta}t, b\vec{r}) = t^{-2x/\theta} \Phi\left(
r t^{-1/\theta}\right) = r^{-2x} \Omega\left( t r^{-\theta}\right)
\end{equation}
where $t$ stands for `temporal' and $\vec{r}$ for `spatial' coordinates,
$x$ is a scaling dimension, $\theta$ the anisotropy exponent (when $t$ 
corresponds to physical time, $\theta=z$ is called the dynamical exponent) and
$\Phi,\Omega$ are scaling functions. Physical realizations of this are numerous,
see \cite{Card96} and references therein. For isotropic critical systems,
$\theta=1$ and the `temporal' variable $t$ becomes just another coordinate. 
It is well-known that in this case, scale invariance (\ref{gl:skala}) with
a constant rescaling factor $b$ can be replaced by the larger group 
of conformal
transformations $b=b(t,\vec{r})$ such that angles are preserved. It turns
out that in the case of one space and one time dimensions, conformal invariance
becomes an important dynamical symmetry from which many physically relevant 
conclusions can be drawn \cite{Bela84}. 

Given the remarkable success of conformal invariance descriptions of 
equilibrium phase transitions, one may wonder whether similar extensions of
scale invariance also exist when $\theta\neq 1$. Indeed, for $\theta=2$ the
analogue of the conformal group is known to be the Schr\"odinger group
\cite{Nied72,Hage72} (and apparently already known to Lie). While applications
of the Schr\"odinger group as dynamical space-time symmetry are known 
\cite{Henk94}, we are interested here in the more general case when 
$\theta\ne 1,2$. We shall first describe the construction of these
{\em local scale transformations}, show that they act as a dynamical
symmetry, then derive the functions $\Phi,\Omega$ and finally comment upon 
some physical applications. For details
we refer the reader to \cite{Henk02}. 

The defining axioms of our notion of {\em local scale invariance} from which
our results will be derived, are as follows (for simplicity, in $d=1$ space
dimensions). 
\begin{enumerate}
\item We seek space-time transformations with infinitesimal generators $X_n$,
such that time undergoes a M\"obius transformation
\BEQ \label{3:Moeb}
t\to t' = \frac{\alpha t + \beta}{\gamma t +\delta} \;\; ; \;\;
\alpha\delta - \beta\gamma =1
\EEQ
and we require that even after the action on the space coordinates is 
included, the commutation relations
\BEQ \label{3:XComm}
\left[ X_n , X_m \right] = (n-m) X_{n+m}
\EEQ
remain valid. This is motivated from the fact that this condition is 
satisfied for both
conformal and Schr\"odinger invariance. 
\item The generator $X_0$ of scale transformations is
\BEQ \label{3:X0gen}
X_0 = - t \partial_t - \frac{1}{\theta} r \partial_r - \frac{x}{\theta}
\EEQ
with a scaling dimension $x$. Similarly, the generator of time translations
is $X_{-1}=-\partial_t$. 
\item Spatial translation invariance is required.
\item Since the Schr\"odinger group acts on wave functions through a projective
representation, generalizations thereof should be expected to occur in the
general case. Such extra terms will be called {\em mass terms}. Similarly, 
extra terms coming from the scaling dimensions should be present. 
\item The generators when applied to a two-point function should yield 
a finite number of independent conditions, i.e. of the form $X_n G=0$.
\end{enumerate}

\noindent
{\bf Proposition 1}: {\it Consider the generators}
\BEQ \label{gl:X}
\!\!\!\!\!\!\!\!\!
X_n = - t^{n+1}\partial_t 
- \sum_{k=0}^{n} \left(\vekz{n+1}{k+1}\right) 
A_{k0}r^{\theta k +1} t^{n-k} \partial_r 
- \sum_{k=0}^{n} \left(\vekz{n+1}{k+1}\right) B_{k0}r^{\theta k} t^{n-k}~~~
\EEQ
{\it where the coefficients $A_{k0}$ and $B_{k0}$ are given by the 
recurrences $A_{n+1,0} = \theta A_{n0} A_{10}$, 
$B_{n+1,0} = \frac{\theta}{n-1}\left( n B_{n0}A_{10} - A_{n0} B_{10}\right)$
for $n\geq 2$ where $A_{00}=1/\theta$, $B_{00}=x/\theta$ and in addition one of 
the following conditions holds: (a) $A_{20}= \theta A_{10}^2$ (b) 
$A_{10}=A_{20}=0$ (c) $A_{20}=B_{20}=0$ (d) $A_{10}=B_{10}=0$. 
These are the most general linear first-order operators in
$\partial_t$ and $\partial_r$ consistent with the above axioms (i) and (ii) 
and which satisfy the commutation relations
$[X_{n},X_{n'}]=(n-n')X_{n+n'}$ for all $n,n'\in\mathbb{Z}$.}

\noindent
Closed but lengthy expressions of the $X_n$ for all $n\in\mathbb{Z}$ are 
known \cite{Henk02}. In order to include space translations, we set
$\theta=2/N$ and use the short-hand 
$X_n = -t^{n+1}\partial_t - a_n \partial_r - b_n$. We then define
\BEQ \label{gl:Y}
Y_m = Y_{k-N/2} = - \frac{2}{N(k+1)}\left(
\frac{\partial a_k(t,r)}{\partial r}\partial_r
+\frac{\partial b_k(t,r)}{\partial r} \right)
\EEQ
where $m=-\frac{N}{2} + k$ and $k$ is an integer. 
Clearly, $Y_{-N/2}=-\partial_r$ generates space translations. 

\noindent {\bf Proposition 2:} 
{\it The generators $X_n$ and $Y_m$ defined in eqs.~(\ref{gl:X},\ref{gl:Y}) 
satisfy the commutation relations}
\BEQ \label{XYComm_II}
\left[ X_n , X_{n'} \right] = (n-n') X_{n+n'} \;\; , \;\;
\left[ X_n , Y_m \right] = \left( n \frac{N}{2} - m \right) Y_{n+m}
\EEQ
{\it in one of the following three cases:
(i) $B_{10}$ arbitrary, $A_{10}=A_{20}=B_{20}=0$ and $N$ arbitrary. 
(ii) $B_{10}$ and $B_{20}$ arbitrary, $A_{10}=A_{20}=0$ and $N=1$.
(iii) $A_{10}$ and $B_{10}$ arbitrary, $A_{20}=A_{10}^2$,
$B_{20}=\frac{3}{2}A_{10} B_{10}$ and $N=2$.}
 
\noindent 
In each case, the generators depend on two free parameters. The physical
interpretation of the free constants $A_{10},A_{20},B_{10},B_{20}$ is still
open. In the cases (ii) and (iii), the generators close into 
a Lie algebra, see \cite{Henk02} for details. For case (i), a closed Lie
algebra exists if $B_{10}=0$. 

Turning to the mass terms, we now restrict to the projective transformations
in time, because we shall only need those in the applications later. 
It is enough to give merely the `special' generator $X_1$ which reads
for $B_{10}=0$ as follows \cite{Henk02}
\BEQ \label{gl:X1}
\!\!
X_1 = -t^2\partial_t - N t r \partial_r - N x t -\alpha r^{2} \partial_t^{N-1}
- \beta r^{2} \partial_r^{2(N-1)/N} - \gamma \partial_r^{2(N-1)/N} r^{2}~~~
\EEQ
where $\alpha,\beta,\gamma$ are free parameters (the cases (ii,iii) of Prop. 2
do not give anything new). Furthermore, it turns out
that the relation $[X_1,Y_{N/2}]=0$ for $N$ integer is only satisfied in one
of the two cases (I) $\beta=\gamma=0$ which we call {\em Type~I} 
and (II) $\alpha=0$ which we call {\em Type~II}. 
In both cases, all generators can be obtained
by repeated commutators of $X_{-1}=-\partial_t$, $Y_{-N/2}=-\partial_r$ and
$X_1$, using (\ref{XYComm_II}). Commutators between two generators $Y_m$ are 
non-trivial and in general only close on certain `physical' states. One might
call such a structure a {\em weak Lie algebra}.  
These results depend on the construction \cite{Henk02} of {\em commuting} 
fractional derivatives satisfying the rules 
$\partial_r^{a+b} =\partial_r^a \partial_r^b$
and $[\partial_r^a,r]=a\partial_r^{a-1}$ (the standard Riemann-Liouville
fractional derivative is not commutative, see e.g. \cite{Hilf00}). 

For $N=1$, the generators of both Type I and Type II reduce to those of the 
Schr\"odinger group. For $N=2$, Type I reproduces the well-known generators of 
$2D$ conformal invariance (without central charge) and Type II gives another
infinite-dimensional group whose Lie algebra is isomorphic to the one of
$2D$ conformal invariance \cite{Henk02}. 

Dynamical symmetries can now be discussed as follows, by calculating the 
commutator of the `Schr\"odinger-operator' $\cal S$ with $X_1$. 
We take $d=1$ and $B_{10}=0$ for simplicity. 

\noindent
{\bf Proposition 3:} {\it The realization of Type I sends any solution 
$\psi(t,{r})$ with scaling dimension $x=1/2-(N-1)/N$ of the 
differential equation}
\BEQ \label{WelleI}
\mathcal{S} \psi(t,\vec{r}) =
\left( -\alpha \partial_t^N + \left(\frac{N}{2}\right)^2
{\partial}_{r}^2 \right) \psi(t,{r}) = 0
\EEQ
{\it into another solution of the same equation.}

\noindent {\bf Proposition 4:} {\it The realization
of Type II sends any solution $\psi(t,r)$ with scaling dimension
$x=(\theta-1)/2+(2-\theta)\gamma/(\beta+\gamma)$ of the differential equation}
\BEQ \label{WelleII}
\mathcal{S} \psi(t,r) =
\left( -(\beta+\gamma) \partial_t + \frac{1}{\theta^2} \partial_r^{\theta}
\right) \psi(t,r) = 0
\EEQ
{\it into another solution of the same equation.}

\noindent 
In both cases, $\cal S$ is a Casimir operator of the `Galilei'-subalgebra
generated from $X_{-1}, Y_{-N/2}$ and the generalized Galilei-transformation
$Y_{-N/2+1}$. The equations (\ref{WelleI},\ref{WelleII}) can be seen as
equations of motion of certain free field theories, where $x$ is the scaling
dimension of that free field $\psi$. These free field theories are
non-local, unless $N$ or $\theta$ are integers, respectively. 

{}From a physical point of view, these wave equations suggest that the 
applications of Types I and II are very different. Indeed, eq.~(\ref{WelleI})
is typical for equilibrium systems with a scaling anisotropy introduced through
competing uniaxial interactions. Paradigmatic cases of this are so-called
Lifshitz points which occur for example in magnetic systems when an
ordered ferromagnetic, a disordered paramagnetic and an incommensurate phase
meet (see \cite{Dieh02} for a recent review). On the other hand, 
eq.~(\ref{WelleII}) is reminiscent of a Langevin equation which may describe
the temporal evolution of a physical system. In any case, causality requirements
can only be met by an evolution equation of first order in $\partial_t$. 

Next, we find the scaling functions $\Phi,\Omega$ in eq.~(\ref{gl:skala}) from
the assumption that $G$ transforms covariantly under local scale 
transformations.

\noindent
{\bf Proposition 5:} {\it Local scale invariance implies that for Type I, 
the function $\Omega(v)$ must satisfy}
\BEQ \label{gl:Omega}
\left( \alpha \partial_v^{N-1} - v^2 \partial_v - N x \right) \Omega(v) = 0
\EEQ
{\it together with the boundary conditions $\Omega(0)=\Omega_0$ and 
$\Omega(v)\sim\Omega_{\infty} v^{-Nx}$ for $v\to\infty$. For Type II, we have}
\BEQ \label{gl:Phi}
\left( \partial_u +\theta(\beta+\gamma)u\partial_u^{2-\theta} 
+2\theta(2-\theta)\gamma\partial_u^{1-\theta}\right)\Phi(u)=0
\EEQ
{\it with the boundary conditions $\Phi(0)=\Phi_0$ and 
$\Phi(u)\sim\Phi_{\infty}u^{-2x}$ for $u\to\infty$.} 

\noindent 
Here $\Omega_{0,\infty}$ and $\Phi_{0,\infty}$ are constants. The ratio
$\beta/\gamma$ turns out to be universal and related to $x$. 
From the linear differential equations (\ref{gl:Omega},\ref{gl:Phi}) 
the scaling functions $\Omega(v)$ and $\Phi(u)$ can be found
explicitly using standard methods \cite{Henk02}. 

Given these explicit results, the idea of local scale invariance can be
tested in specific models. Indeed, the predictions for $\Omega(v)$ coming from
Type I with $N=4$ nicely agree with cluster Monte Carlo data for the 
spin-spin and energy-energy correlators of the $3D$ ANNNI model at its
Lifshitz point \cite{Henk02,Plei01}. On the other hand, the predictions of 
Type II have been tested extensively in the context of ageing ferromagnetic 
spin systems to which we turn now. 

Consider a ferromagnetic spin system (e.g. an Ising model) prepared in a
high-temperature initial state and then quenched to some temperature $T$ at or
below the critical temperature $T_c$. Then the system is left to evolve freely 
(for recent reviews, see \cite{Cate00,Godr02}). 
It turns out that clusters of a typical time-dependent size $L(t)\sim t^{1/z}$
form and grow, where $z$ is the dynamical exponent. Furthermore, two-time
observables such as the response function $R(t,s;\vec{r}-\vec{r}')=
\delta\langle\sigma_{\vec{r}}(t)\rangle/\delta h_{\vec{r}'}(s)$ depend on 
{\em both} $t$ and $s$, where
$\sigma_{\vec{r}}$ is a spin variable and $h_{\vec{r}'}$ the conjugate magnetic
field. This breaking of time-translation
invariance is called {\em ageing}. We are mainly interested in the
autoresponse function $R(t,s)=R(t,s;\vec{0})$. One finds a dynamic scaling 
behaviour $R(t,s)\sim s^{-1-a} f_R(t/s)$ with $f_R(x)\sim x^{-\lambda_R/z}$
for $x\gg 1$ and where $\lambda_R$ and $a$ are exponents to be determined.

In order to apply local scale invariance 
to this problem, we must take into account that time translation invariance does
{\em not} hold. The simplest way to do this is to remark that the 
Type~II-subalgebra spanned by $X_0, X_1$ and the $Y_m$ leaves the initial 
line $t=0$ invariant, see (\ref{gl:X1}). Therefore the autoresponse function
$R(t,s)$ is fixed by the two covariance conditions $X_0 R = X_1 R=0$. Solving
these differential equations and comparing with the above scaling forms
leads to \cite{Henk02}
\BEQ \label{gl:R}
R(t,s) = r_0 \left( t/s \right)^{1+a-\lambda_R/z} \left( t-s\right)^{-1-a}
\;\; , \;\; t > s~~~~
\EEQ  
where $r_0$ is a normalization constant. Therefore the functional form of $R$ 
is completely fixed once the exponents
$a$ and $\lambda_R/z$ are known. Similarly, the spatio-temporal
response $R(t,s;\vec{r})=R(t,s) \Phi\left(r (t-s)^{-1/z}\right)$, with the
scaling function $\Phi(u)$ determined by (\ref{gl:Phi}).

The prediction (\ref{gl:R}) has been confirmed recently in several physically
distinct systems undergoing ageing, see \cite{Henk01,Cann01,Cala02,Pico02} and
references therein. These confirmations (which go beyond free field theory) 
suggest that (\ref{gl:R}) should hold
independently of (i) the value of the dynamical exponent $z$ (ii) the spatial
dimensionality $d>1$ (iii) the numbers of components of the order parameter and
the global symmetry group (iv) the spatial range of the interactions (v) the
presence of spatially long-range initial correlations (vi) the value of the
temperature $T$ (vii) the presence of weak disorder. Evidently, additional model
studies are called for to test this conjecture further.

Summarizing, we have shown that local scale transformations exist for any 
$\theta$, act as dynamical symmetries of certain non-local free field theories
and appear to be realized as space-time symmetries in some strongly anisotropic 
critical systems of physical interest.

\end{document}